%% file: main.tex
\documentclass{article}
\usepackage{spconf,amsmath,amssymb,graphicx,bm}
\usepackage{booktabs}
\usepackage[export]{adjustbox}
\usepackage{microtype}
\usepackage{xcolor}
\usepackage{makecell}
\usepackage{multirow}

\usepackage{tikz}
\usetikzlibrary{arrows.meta, calc}
\usepackage{pgfplots}
\usepgfplotslibrary{groupplots}
\pgfplotsset{compat=1.18}

\usepackage[backend=biber, bibencoding=utf8,
            style=ieee,
            maxbibnames=6,
            doi=false,
            url=false,
            isbn=false]{biblatex}
\AtEveryBibitem{%
  \ifentrytype{inproceedings}{%
    \clearfield{pages}%
    \clearlist{publisher}%
    \clearlist{organization}%
    \clearlist{location}}{}%
  \ifentrytype{misc}{%
    \clearfield{eprintclass}{} }}

\usepackage[hidelinks]{hyperref}

\usepackage[nolist]{acronym}
\input{acronyms.tex}

\newcommand\blfootnote[1]{%
  \begingroup
  \renewcommand\thefootnote{}\footnote{#1}%
  \addtocounter{footnote}{-1}%
  \endgroup
}

\title{Perceptual Quality Loss or Loss of Perceptual Quality?}
\name{Danilo {de Oliveira}, Tal Peer, Maur\'icio {do V. M. da Costa}, Timo Gerkmann}
\address{Signal Processing, University of Hamburg, Germany}
\begin{document}
\ninept
\maketitle
\begin{abstract}
    Contemporary deep speech enhancement (SE) models are often trained with specific auxiliary terms in the loss function as a way to improve their performance in terms of perceptual metrics. Nevertheless, a higher score on a perceptual metric does not necessarily correlate with an improved listening experience. Through objective and subjective experiments, we assess the performance of SE models trained with two different types of auxiliary PESQ loss terms. The numerical evaluation on a suite of standard metrics suggests that, while models optimized for PESQ naturally obtain higher PESQ scores in the test set, for most other metrics the scores do not significantly change. In some cases, the PESQ loss even results in worse PESQ scores on mismatched data. A formal listening experiment reveals that the models without a PESQ loss were generally preferred over models that include it, across all settings. Finally, we analyze the relative importance of PESQ in the composite metrics CSIG, CBAK and COVL, and find that PESQ dominates all of them. Our study highlights the perils of over-reliance on PESQ and stresses the importance of a complete evaluation procedure for SE.
\end{abstract}

\section{INTRODUCTION}
\label{sec:intro}

Evaluation of \ac{SE} models is a rich research topic; due to speech quality being highly subjective, reliable \ac{SQA} is a challenging task~\cite{loizou2013speech}. There exists a vast literature of listening evaluation procedures, formalized in \ac{ITU} recommendations \cite{itut1996p800, itut2003p835}. Given the time-consuming nature of such listening experiments, researchers have long devised instrumental measures of speech quality that provide direct feedback, enabling faster iteration in the development of noise suppression algorithms. \Acp{DNN} have fostered the development of non-intrusive (reference-free) metrics~\cite{ragano2024scoreq, stahl2025distillation}. However, metrics continue to face challenges inherent to the task, e.g. generalization~\cite{cooper2022generalization} and noise in the subjective ratings~\cite{cumlin2026rhoperfect}.

Perceptual quality metrics can be incorporated into \ac{SE} model optimization, with the goal of improving perceptual quality in the model's outputs. In cases where the metric is differentiable, this can be done in a straightforward manner; nonetheless, non-differentiable metrics can also be included, e.g. via a metric predictor that acts as a differentiable surrogate~\cite{fu2019metricgan}. A commonly used metric in this context is the \ac{PESQ}~\cite{rix2001pesq}, an intrusive (reference-based) perceptual measure of relatively lightweight computation whose implementation is readily available online.\footnote{\url{https://github.com/ludlows/PESQ}} However, as metrics are imperfect proxies of human perception, optimization of \acp{DNN} based on them is prone to unintended artifacts or even complete failure. In \cite{deoliveira2024pesqetarian}, it was shown that high \ac{PESQ} scores do not necessarily correlate with high perceived quality. The PESQetarian, a model optimized with a \ac{PESQ}-focused objective, yields high \ac{PESQ} scores while simultaneously performing very poorly in a human listening experiment. Furthermore, it was shown that high \ac{PESQ} scores can be obtained even when distortions are purposefully induced.

The PESQetarian approach in~\cite{deoliveira2024pesqetarian} was designed to demonstrate the inherent problems arising from optimizing and evaluating a model using the same metric. For this, the extreme case of optimizing exclusively for \ac{PESQ} was considered. In practice, however, metric losses are typically not used exclusively, but are rather combined with regular distance-based losses (e.g. MSE or SI-SDR)~\cite{cao2022cmgan, kim2019endtoend}. Yet, this setting is still susceptible to Goodhart's law, which states that ``when a measure becomes a target, it ceases to be a good measure''~\cite{strathern1997improving}. An \ac{SE} model optimized with a metric loss should naturally obtain higher performance in that metric. As a result, in terms of this metric, it ceases to be directly comparable to models that have not been optimized for that specific metric. Furthermore, even if constrained by other loss terms, the metric's imperfect perceptual modeling can still be detrimental to performance on other metrics~\cite{richter2025investigating}. 

In this paper, we thoroughly investigate the effect of the \ac{PESQ} loss on \ac{SE} performance according to a suite of instrumental metrics, accompanied by a formal listening experiment. Our investigation includes two different versions of a \ac{PESQ} loss, as featured in two representative \ac{SE} systems~\cite{richter2025investigating, chao2024investigation}. Additionally, we analyze the metrics that are included in a widely used evaluation script of composite measures~\cite{hu2006evaluation}. Our results emphasize the importance of an evaluation process based on a comprehensive panel of metrics and datasets, ideally including data collected from a significantly large pool of human subjects. Moreover, they demonstrate the pitfalls of metric optimization: even as an auxiliary objective, perceptual metrics may not induce any meaningful improvement or even be detrimental to \ac{SE} performance. Audio samples and an evaluation script are available on our project webpage.\footnote{\url{https://uhh.de/inf-sp-pesqloss}}
\vspace{-.1cm}
\section{PESQ-BASED LOSS FUNCTIONS}
\label{sec:method}

\begin{figure}[t]
  \centering
  \includegraphics[width=.98\linewidth]{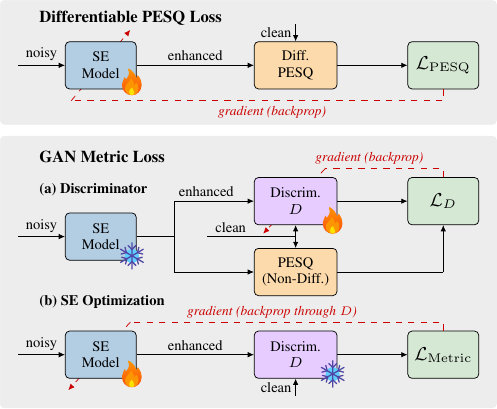}
\caption{Differentiable PESQ loss and GAN metric loss frameworks.}
\label{fig:lossdiagram}
\vspace{-0.3cm}
\end{figure}

\ac{PESQ} is a standard reference-based objective metric for \ac{SQA}. It predicts the \ac{MOS}~\cite{itut1996p800} that would be assigned to a potentially degraded audio sample. Designed for telecommunication settings, \ac{PESQ} was standardized by the \ac{ITU} in 2001 as recommendation ITU-T P.862~\cite{itut2001p862} and superseded in 2011 by \ac{POLQA}~\cite{beerends2013perceptual}, in recommendation ITU-T P.863~\cite{itut2011p863}. Nevertheless, it remains a commonly used method and a standard benchmark for instrumental evaluation of \ac{SE} algorithms.

\ac{PESQ} comprises time- and level-alignment stages for normalization of reference and degraded signals, followed by a perceptually-motivated auditory transform. In the transformed time-frequency representation, disturbances w.r.t. the reference signal are computed, aggregated, and mapped into the \ac{MOS} scale, i.e. it ranges from 1 to 5, with higher scores representing better perceptual quality.

\subsection{Differentiable PESQ Loss}\label{subsec:diffpesq}

Based on \cite{martindonas2018deeplearning, kim2019endtoend}, the \texttt{torch-pesq}\footnote{\url{https://github.com/audiolabs/torch-pesq}} package adapts the \ac{PESQ} computation steps in a differentiable implementation, as illustrated in Fig.~\ref{fig:lossdiagram} (top). Contrasting with the original implementation, time-alignment operations are skipped, and level alignment is performed via IIR filtering. The authors of the implementation have reported that optimizing  \ac{SE} models for the \ac{PESQ} loss alone does not yield good results, and suggested that an \ac{SI-SDR}~\cite{leroux2019sdr} loss term should be used in conjunction with it.\\

\noindent\textbf{SB-SGMSE+} consists of a diffusion-based model employing a \ac{SB}~\cite{chen2022likelihood} to describe the transformation path between the distributions of noisy and clean speech signals~\cite{richter2025investigating}. The framework of \acp{SB} allows for the incorporation of data prediction losses, e.g. an $L_1$ loss term in the time domain~\cite{jukic2024schroedinger}. In \cite{richter2025investigating}, the authors propose including a differentiable \ac{PESQ} loss term, weighted by a hyperparameter $\alpha_P$. In this manuscript, models M5, M7, and M6 are evaluated, with $\alpha_P$ assuming the values of 0 (no PESQ), 0.0005 (PESQ-mid), and 0.001 (PESQ-high), respectively. The checkpoints made available by the authors are used.\footnote{\url{https://github.com/sp-uhh/sgmse}}

\subsection{GAN Metric Loss}\label{subsec:metricgan}

Pioneered by MetricGAN~\cite{fu2019metricgan}, the \ac{GAN}-style~\cite{goodfellow2014generative} training loss term treats the target metric as a black box and trains a \ac{DNN} to act as a differentiable surrogate. Despite being metric-agnostic, \ac{PESQ} was the metric of choice in many subsequent \ac{SE} works~\cite{xu2022deep, cao2022cmgan, lu2023mpsenet}.

Figure~\ref{fig:lossdiagram} (bottom) depicts the GAN metric loss term described in the following. The discriminator $D$ takes a reference-degraded pair of magnitude spectrograms as inputs and outputs a prediction of \ac{PESQ} rescaled in the range $[0, 1]$. The discriminator loss is
\begin{equation}\begin{split}
    \mathcal{L}_D &= \mathbb{E}_{\bm{X_{m}}}[\|D(\bm{X_{m}}, \bm{X_{m}})-1\|_{2}^{2}]\\
    &+\mathbb{E}_{\bm{X_{m}},\bm{\hat{X}_{m}}}[\|D(\bm{X_{m}},\bm{\hat{X}_{m}})-Q_{\rm \ac{PESQ}}\|_{2}^{2}],
\end{split}\end{equation}
where $\bm{X_{m}}$ is the clean magnitude spectrogram, $\bm{\hat{X}_{m}}$ is the enhanced magnitude spectrogram estimated by the \ac{SE} model in the current training iteration, and $Q_{\rm \ac{PESQ}}$ is the actual \ac{PESQ} score rescaled. This means that the metric discriminator is trained to estimate \ac{PESQ}, encouraging the assignment of the highest score to identical signal pairs. In turn, the generator metric loss term  $\mathcal{L}_{\rm Metric}$ encourages the \ac{SE} model to produce outputs with the highest perceptual score, as judged by the discriminator:
\begin{equation}
    \mathcal{L}_{\rm Metric} = \mathbb{E}_{\bm{X_{m}},\bm{\hat{X}_{m}}}[\|D(\bm{X_{m}},\bm{\hat{X}_{m}})-1\|_{2}^{2}].
\end{equation}

\noindent\textbf{SEMamba}: Building on the dual magnitude/phase decoder design of MP-SENet~\cite{lu2023mpsenet}, the authors of~\cite{chao2024investigation} investigated replacing the two-stage Conformer blocks in the backbone with Mamba~\cite{gu2024mamba} blocks. The resulting model was named SEMamba-advanced, which is referred to as SEMamba in the remainder of this paper. It combines a suite of loss terms based on waveform amplitude, complex \ac{STFT} coefficients, \ac{STFT} consistency, and a \ac{GAN}-based metric loss, as described above. The official code provided by the authors\footnote{\url{https://github.com/RoyChao19477/SEMamba}} is used, retraining with batch size 12 and learning rate 0.001. We train two versions: one excluding the \ac{GAN} loss from the aforementioned suite of loss terms  (SEMamba w/o metric loss), and one with the original loss (SEMamba w/ metric loss), for a direct comparison.

\section{EXPERIMENTS}\label{sec:experiments}

\begin{table*}[ht]
    \caption{Results of SB-SGMSE+ on the matched VB-DMD and the mismatched EARS-WHAM datasets. Best results highlighted in bold font, alongside other methods whose difference is statistically insignificant according to a paired Wilcoxon signed-rank test. The upper set corresponds to CMGAN's evaluation script, with the updated ESTOI metric. POLQA$^\ast$ indicates only a subset of files produced valid scores.}
    \label{tab:sbsgmse}
    \centering
    \begin{adjustbox}{width=\linewidth}
    \addtolength{\tabcolsep}{-0.4em}
    \begin{tabular}{lcccccccc}
        \toprule
         & \multicolumn{4}{c}{VB-DMD (Matched)} & \multicolumn{4}{c}{EARS-WHAM (Mismatched)} \\
        \cmidrule(lr){2-5}\cmidrule(lr){6-9}
        Metric & Noisy & \makecell{SB-SGMSE+\\M5 (no PESQ)} & \makecell{SB-SGMSE+\\M7 (PESQ-mid)} & \makecell{SB-SGMSE+\\M6 (PESQ-high)} & Noisy & \makecell{SB-SGMSE+\\M5 (no PESQ)} & \makecell{SB-SGMSE+\\M7 (PESQ-mid)} & \makecell{SB-SGMSE+\\M6 (PESQ-high)} \\
        \midrule
        PESQ & 1.97 {\scriptsize$\pm$ 0.75} & 2.91 {\scriptsize$\pm$ 0.75} & 3.56 {\scriptsize$\pm$ 0.65} & \textbf{3.73 {\scriptsize$\pm$ 0.56}} & 1.24 {\scriptsize$\pm$ 0.21} & 1.85 {\scriptsize$\pm$ 0.62} & 2.26 {\scriptsize$\pm$ 0.70} & \textbf{2.32 {\scriptsize$\pm$ 0.68}} \\
        CSIG & 3.49 {\scriptsize$\pm$ 0.81} & 4.03 {\scriptsize$\pm$ 0.83} & 4.63 {\scriptsize$\pm$ 0.52} & \textbf{4.69 {\scriptsize$\pm$ 0.45}} & 2.75 {\scriptsize$\pm$ 0.53} & 2.72 {\scriptsize$\pm$ 0.84} & 3.31 {\scriptsize$\pm$ 0.80} & \textbf{3.36 {\scriptsize$\pm$ 0.77}} \\
        CBAK & 2.55 {\scriptsize$\pm$ 0.64} & 3.64 {\scriptsize$\pm$ 0.53} & \textbf{3.79 {\scriptsize$\pm$ 0.43}} & 3.73 {\scriptsize$\pm$ 0.38} & 2.10 {\scriptsize$\pm$ 0.38} & 2.83 {\scriptsize$\pm$ 0.52} & \textbf{2.93 {\scriptsize$\pm$ 0.51}} & 2.84 {\scriptsize$\pm$ 0.48} \\
        COVL & 2.74 {\scriptsize$\pm$ 0.79} & 3.53 {\scriptsize$\pm$ 0.80} & 4.21 {\scriptsize$\pm$ 0.64} & \textbf{4.33 {\scriptsize$\pm$ 0.56}} & 2.02 {\scriptsize$\pm$ 0.37} & 2.33 {\scriptsize$\pm$ 0.72} & 2.84 {\scriptsize$\pm$ 0.75} & \textbf{2.88 {\scriptsize$\pm$ 0.73}} \\
        ESTOI & 0.79 {\scriptsize$\pm$ 0.15} & \textbf{0.88 {\scriptsize$\pm$ 0.10}} & 0.87 {\scriptsize$\pm$ 0.09} & 0.86 {\scriptsize$\pm$ 0.09} & 0.64 {\scriptsize$\pm$ 0.17} & \textbf{0.77 {\scriptsize$\pm$ 0.17}} & 0.76 {\scriptsize$\pm$ 0.17} & 0.74 {\scriptsize$\pm$ 0.18} \\
        segSNR [dB] & 1.68 {\scriptsize$\pm$ 4.44} & \textbf{10.18 {\scriptsize$\pm$ 3.23}} & 7.59 {\scriptsize$\pm$ 2.71} & 5.53 {\scriptsize$\pm$ 2.39} & -0.80 {\scriptsize$\pm$ 3.98} & \textbf{5.28 {\scriptsize$\pm$ 3.96}} & 3.88 {\scriptsize$\pm$ 3.28} & 2.06 {\scriptsize$\pm$ 2.95} \\
        \midrule
        SI-SDR [dB] & 8.44 {\scriptsize$\pm$ 5.62} & \textbf{19.43 {\scriptsize$\pm$ 3.48}} & 13.21 {\scriptsize$\pm$ 2.87} & 7.70 {\scriptsize$\pm$ 2.86} & 5.36 {\scriptsize$\pm$ 5.91} & \textbf{12.67 {\scriptsize$\pm$ 5.65}} & 7.90 {\scriptsize$\pm$ 3.77} & 3.23 {\scriptsize$\pm$ 3.40} \\
        POLQA$^\ast$ & 3.23 {\scriptsize$\pm$ 0.81} & 4.32 {\scriptsize$\pm$ 0.54} & \textbf{4.40 {\scriptsize$\pm$ 0.47}} & 4.22 {\scriptsize$\pm$ 0.55} & 2.03 {\scriptsize$\pm$ 0.51} & 2.98 {\scriptsize$\pm$ 1.00} & \textbf{3.13 {\scriptsize$\pm$ 1.04}} & 2.77 {\scriptsize$\pm$ 1.01} \\
        DistillMOS & 3.64 {\scriptsize$\pm$ 0.37} & \textbf{3.90 {\scriptsize$\pm$ 0.31}} & 3.78 {\scriptsize$\pm$ 0.34} & 3.73 {\scriptsize$\pm$ 0.37} & 2.58 {\scriptsize$\pm$ 0.60} & \textbf{4.16 {\scriptsize$\pm$ 0.58}} & 4.11 {\scriptsize$\pm$ 0.64} & 3.79 {\scriptsize$\pm$ 0.78} \\
        SCOREQ (NR) & 3.32 {\scriptsize$\pm$ 0.64} & 4.34 {\scriptsize$\pm$ 0.33} & \textbf{4.39 {\scriptsize$\pm$ 0.30}} & 4.27 {\scriptsize$\pm$ 0.44} & 2.13 {\scriptsize$\pm$ 0.58} & 3.12 {\scriptsize$\pm$ 0.91} & \textbf{3.17 {\scriptsize$\pm$ 0.87}} & 2.82 {\scriptsize$\pm$ 0.88} \\
        SCOREQ (REF) $\downarrow$ & 0.76 {\scriptsize$\pm$ 0.28} & \textbf{0.27 {\scriptsize$\pm$ 0.14}} & \textbf{0.26 {\scriptsize$\pm$ 0.13}} & 0.32 {\scriptsize$\pm$ 0.17} & 1.13 {\scriptsize$\pm$ 0.21} & 0.62 {\scriptsize$\pm$ 0.33} & \textbf{0.60 {\scriptsize$\pm$ 0.30}} & 0.73 {\scriptsize$\pm$ 0.30} \\
        WAcc {\scriptsize QuartzNet 15x5} [\%] & 91.83 {\scriptsize$\pm$ 16.06} & \textbf{93.68 {\scriptsize$\pm$ 14.04}} & \textbf{93.49 {\scriptsize$\pm$ 13.96}} & 92.67 {\scriptsize$\pm$ 15.27} & 67.22 {\scriptsize$\pm$ 29.35} & \textbf{70.46 {\scriptsize$\pm$ 27.64}} & \textbf{70.04 {\scriptsize$\pm$ 27.72}} & 67.99 {\scriptsize$\pm$ 29.05} \\
        WAcc {\scriptsize Parakeet CTC 0.6B} [\%] & 97.23 {\scriptsize$\pm$ 10.94} & \textbf{97.99 {\scriptsize$\pm$ 7.35}} & \textbf{97.94 {\scriptsize$\pm$ 9.32}} & \textbf{97.85 {\scriptsize$\pm$ 9.66}} & 90.60 {\scriptsize$\pm$ 16.84} & \textbf{86.38 {\scriptsize$\pm$ 20.92}} & 84.61 {\scriptsize$\pm$ 22.71} & 83.28 {\scriptsize$\pm$ 23.43} \\
        \bottomrule
    \end{tabular}
    \end{adjustbox}
    \vspace{-0.3cm}
\end{table*}

\begin{table*}[ht]
    \caption{Results of SEMamba on the matched VB-DMD and the mismatched EARS-WHAM datasets. Best results highlighted in bold font, alongside other methods whose difference is statistically insignificant according to a paired Wilcoxon signed-rank test. The upper set corresponds to CMGAN's evaluation script, with the updated ESTOI metric. POLQA$^\ast$ indicates only a subset of files produced valid scores.}
    \label{tab:semamba}
    \centering
    \begin{adjustbox}{width=.8\linewidth}
    \begin{tabular}{lcccccc}
        \toprule
         & \multicolumn{3}{c}{VB-DMD (Matched)} & \multicolumn{3}{c}{EARS-WHAM (Mismatched)} \\
        \cmidrule(lr){2-4}\cmidrule(lr){5-7}
        Metric & Noisy & \makecell{SEMamba\\w/o metric loss} & \makecell{SEMamba\\w/ metric loss} & Noisy & \makecell{SEMamba\\w/o metric loss} & \makecell{SEMamba\\w/ metric loss} \\
        \midrule
        PESQ Discriminator $D$ & 2.45 {\scriptsize$\pm$ 0.58} & 3.26 {\scriptsize$\pm$ 0.58} & \textbf{3.34 {\scriptsize$\pm$ 0.56}} & 1.61 {\scriptsize$\pm$ 0.15} & \textbf{1.78 {\scriptsize$\pm$ 0.57}} & 1.75 {\scriptsize$\pm$ 0.57} \\
        \midrule
        PESQ & 1.97 {\scriptsize$\pm$ 0.75} & 3.35 {\scriptsize$\pm$ 0.63} & \textbf{3.42 {\scriptsize$\pm$ 0.60}} & 1.24 {\scriptsize$\pm$ 0.21} & \textbf{2.21 {\scriptsize$\pm$ 0.70}} & 2.16 {\scriptsize$\pm$ 0.72} \\
        CSIG & 3.49 {\scriptsize$\pm$ 0.81} & 4.71 {\scriptsize$\pm$ 0.43} & \textbf{4.72 {\scriptsize$\pm$ 0.42}} & 2.75 {\scriptsize$\pm$ 0.53} & \textbf{3.73 {\scriptsize$\pm$ 0.67}} & 3.65 {\scriptsize$\pm$ 0.72} \\
        CBAK & 2.55 {\scriptsize$\pm$ 0.64} & 3.88 {\scriptsize$\pm$ 0.47} & \textbf{3.90 {\scriptsize$\pm$ 0.47}} & 2.10 {\scriptsize$\pm$ 0.38} & \textbf{2.97 {\scriptsize$\pm$ 0.56}} & 2.89 {\scriptsize$\pm$ 0.56} \\
        COVL & 2.74 {\scriptsize$\pm$ 0.79} & 4.13 {\scriptsize$\pm$ 0.58} & \textbf{4.17 {\scriptsize$\pm$ 0.55}} & 2.02 {\scriptsize$\pm$ 0.37} & \textbf{3.02 {\scriptsize$\pm$ 0.69}} & 2.95 {\scriptsize$\pm$ 0.72} \\
        ESTOI & 0.79 {\scriptsize$\pm$ 0.15} & \textbf{0.89 {\scriptsize$\pm$ 0.08}} & \textbf{0.89 {\scriptsize$\pm$ 0.08}} & 0.64 {\scriptsize$\pm$ 0.17} & \textbf{0.78 {\scriptsize$\pm$ 0.17}} & 0.76 {\scriptsize$\pm$ 0.18} \\
        segSNR [dB] & 1.68 {\scriptsize$\pm$ 4.44} & \textbf{10.69 {\scriptsize$\pm$ 3.25}} & \textbf{10.46 {\scriptsize$\pm$ 3.57}} & -0.80 {\scriptsize$\pm$ 3.98} & \textbf{4.72 {\scriptsize$\pm$ 4.08}} & 3.85 {\scriptsize$\pm$ 3.98} \\
        \midrule
        SI-SDR [dB] & 8.44 {\scriptsize$\pm$ 5.62} & \textbf{20.11 {\scriptsize$\pm$ 3.40}} & 19.70 {\scriptsize$\pm$ 3.81} & 5.36 {\scriptsize$\pm$ 5.91} & \textbf{12.06 {\scriptsize$\pm$ 5.67}} & 9.96 {\scriptsize$\pm$ 5.21} \\
        POLQA$^\ast$ & 3.23 {\scriptsize$\pm$ 0.81} & \textbf{4.44 {\scriptsize$\pm$ 0.41}} & 4.43 {\scriptsize$\pm$ 0.41} & 2.03 {\scriptsize$\pm$ 0.51} & \textbf{3.26 {\scriptsize$\pm$ 0.99}} & 3.14 {\scriptsize$\pm$ 1.02} \\
        DistillMOS & 3.64 {\scriptsize$\pm$ 0.37} & \textbf{3.91 {\scriptsize$\pm$ 0.31}} & 3.90 {\scriptsize$\pm$ 0.32} & 2.58 {\scriptsize$\pm$ 0.60} & 3.77 {\scriptsize$\pm$ 0.77} & \textbf{3.82 {\scriptsize$\pm$ 0.81}} \\
        SCOREQ (NR) & 3.32 {\scriptsize$\pm$ 0.64} & \textbf{4.42 {\scriptsize$\pm$ 0.25}} & 4.41 {\scriptsize$\pm$ 0.27} & 2.13 {\scriptsize$\pm$ 0.58} & \textbf{3.20 {\scriptsize$\pm$ 0.90}} & \textbf{3.20 {\scriptsize$\pm$ 0.88}} \\
        SCOREQ (REF) $\downarrow$ & 0.76 {\scriptsize$\pm$ 0.28} & \textbf{0.18 {\scriptsize$\pm$ 0.10}} & \textbf{0.18 {\scriptsize$\pm$ 0.10}} & 1.13 {\scriptsize$\pm$ 0.21} & \textbf{0.58 {\scriptsize$\pm$ 0.34}} & \textbf{0.57 {\scriptsize$\pm$ 0.33}} \\
        WAcc {\scriptsize QuartzNet 15x5} [\%] & 91.83 {\scriptsize$\pm$ 16.06} & \textbf{94.30 {\scriptsize$\pm$ 12.88}} & \textbf{94.45 {\scriptsize$\pm$ 13.98}} & 67.22 {\scriptsize$\pm$ 29.35} & \textbf{74.03 {\scriptsize$\pm$ 25.34}} & 72.66 {\scriptsize$\pm$ 26.30} \\
        WAcc {\scriptsize Parakeet CTC 0.6B} [\%] & 97.23 {\scriptsize$\pm$ 10.94} & \textbf{98.31 {\scriptsize$\pm$ 8.10}} & \textbf{98.24 {\scriptsize$\pm$ 9.57}} & 91.60 {\scriptsize$\pm$ 16.84} & \textbf{87.64 {\scriptsize$\pm$ 19.35}} & 86.11 {\scriptsize$\pm$ 21.02} \\
        \bottomrule
    \end{tabular}
    \end{adjustbox}
    \vspace{-0.3cm}
\end{table*}

\subsection{Instrumental Metrics}\label{subsec:metrics}

Aiming at obtaining a broad view of the effect of \ac{PESQ} optimization over \ac{SE} performance, a variety of metrics is employed, both reference-based and reference-free, as listed below.

\vspace{.1cm}

\noindent\textbf{Composite metrics (CSIG, CBAK, COVL)}~\cite{hu2006evaluation} are combinations of other instrumental measures, fit through linear regression to correlate with listening scores from recommendation ITU-T P.835~\cite{itut2003p835}.

\vspace{.1cm}

\noindent\textbf{ESTOI}~\cite{jensen2016algorithm} is an intrusive metric for speech intelligibility, based on the spectral correlation between temporal envelopes of reference and degraded signals. It is expected to have a monotonic, non-linear, and dataset-dependent relation with speech intelligibility.

\vspace{.1cm}

\noindent\textbf{segSNR} is the average \ac{SNR} computed over windowed segments.

\vspace{.1cm}

\noindent\textbf{SI-SDR} measures the level of distortion (residual noise, artifacts and, in source separation, interfering sources) between degraded and reference signals in a scale-invariant manner.

\vspace{.1cm}

\noindent\textbf{POLQA}~\cite{beerends2013perceptual} is the successor of \ac{PESQ}. The approach is similar to its predecessor, with a series of improvements aimed at modern telecommunication systems, including support for full-band audio.

\vspace{.1cm}

\noindent\textbf{WAcc} is the word accuracy, which measures the performance of \ac{ASR} models on the enhanced audio. It is defined as $1-\text{WER}$, where WER is the word error rate, i.e. the edit distance on the text transcriptions and thus a content-intrusive metric. In \cite{deoliveira2026toogood}, it was shown that systems with language modeling capabilities can overcome acoustic artifacts. For a focus on acoustics, we therefore make exclusive use \ac{ASR} models based on \ac{CTC}~\cite{graves2006connectionist}. The text normalization pipeline corresponds to the \texttt{wer\_standardize} pipeline from the \texttt{jiwer} package, with additional punctuation removal, expansion of informal contractions, and number-to-word conversion, as in~\cite{deoliveira2026toogood}. 

\vspace{.1cm}

\noindent\textbf{SCOREQ} \cite{ragano2024scoreq} is a \ac{DNN}-based model of \ac{SQA} whose training involves contrastive learning for structuring the representation space according to \ac{MOS}. It has two modes of operation: reference-based (REF), where the distance between embeddings of clean and degraded audio signals is calculated; and reference-free (NR), where the embeddings are mapped into the \ac{MOS} scale via fine-tuning.

\vspace{.1cm}

\noindent\textbf{DistillMOS}~\cite{stahl2025distillation} is a convolutional model trained via distillation of a large self-supervised teacher model, with human-labeled data as targets, as well as unlabeled data ``pseudo-labeled'' by the teacher. 

\subsection{Data}\label{subsec:data}

Our analysis considers two \ac{SE} benchmarks. For a matched setting, the \ac{VB-DMD} test set~\cite{valentini2016investigating} is used, corresponding to the training set used to optimize the \ac{SE} models. This test set contains 824 samples with input \acp{SNR} in the set $\{2.5, 5.0, 7.5, \dots, 17.5\}$~dB. For an investigation of generalization to a mismatched, more challenging setting, the EARS-WHAM v2 test set~\cite{richter2024ears} is used. It contains 886 mixtures with \ac{SNR} continuously distributed in the range $[-2.5, 17.5]$~dB. Both datasets are down-sampled to 16~kHz. 

\vspace{-.1cm}
\subsection{Listening Experiment}\label{subsec:experiment}
\vspace{.1cm}

A formal listening experiment was conducted to assess the effect that objective perceptual loss optimization terms have on the subjective perceived quality of enhanced speech signals. The experiment was designed as a blind preference test, where participants were presented with two stimuli at a time and asked to choose the best-sounding one in terms of overall perceived quality. For SEMamba, the choice was between the default model with all loss terms and one without the $\mathcal{L}_{\rm Metric}$ term. For SB-SGMSE+, triplets of samples from models M5, M6, and M7 were presented to participants in pairwise comparisons, distributed randomly throughout the experiment. Thirty samples from each test set were selected. Stimuli were assigned randomly to participants, with the constraints of stratified sampling w.r.t. input \ac{SNR}. For \ac{VB-DMD}, whose sentences are generally short, files shorter than 3 seconds are filtered out. In the case of EARS-WHAM, where samples are longer, segments of approximately 6 seconds are cropped, avoiding mid-sentence crops through the use of a \ac{VAD} on the corresponding clean audio files. All stimuli were normalized to -23~LUFS target loudness. 

The experiment was conducted in a sound-proofed listening booth, using Beyerdynamic DT 770 PRO headphones and a fixed volume level across all participants. Twelve audio experts participated in the experiment, each rating 56 pairs of speech signals. Of these, eight pairs were screening (dummy) tests. Three dummy tests were presented in the beginning for training purposes, while the other five were distributed randomly throughout the experiment to probe for continued attention of the participants to the task. Each stimulus was evaluated by three different participants.

\subsection{Composite Metric Analysis}\label{subsec:composite}

The set of metrics containing \ac{PESQ}, CSIG, CBAK, COVL, \ac{segSNR} and ESTOI's predecessor STOI was used in \cite{cao2022cmgan} and has become a popular choice of evaluation procedure. We conduct an analysis of the importance of each component involved in the computation of CSIG, CBAK, and COVL; these are \ac{PESQ}, log-likelihood ratio, \ac{WSS} distance~\cite{klatt1982prediction} and \ac{segSNR}. Since each component operates in a different range, each metric's share of the total variance in the linear part of the composite is computed. The empirical statistics are calculated from the data of \ac{VB-DMD} and EARS-WHAM, noisy and enhanced by the models considered in this paper.

\begin{figure}[t]
    \centering
    \includegraphics[width=\linewidth]{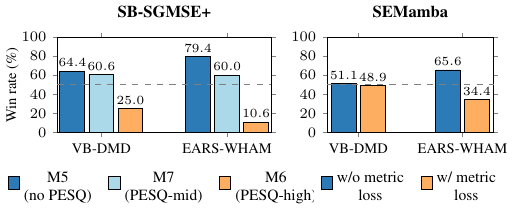}
    \vspace{-0.7cm}
    \caption{Listening experiment win rates (\%) by model and dataset.}
    \label{fig:winrate}
    \vspace{-0.3cm}
\end{figure}

\vspace{-.2cm}
\section{RESULTS}
\label{sec:results}

Tables~\ref{tab:sbsgmse} and~\ref{tab:semamba} present evaluation results in terms of the instrumental metrics described in Sec.~\ref{subsec:metrics}. Due to minimum speech length constraints, POLQA could only be computed on 56\% and 98\% of the \ac{VB-DMD} and EARS-WHAM test samples, respectively.
Comparisons are made within each test set. For SB-SGMSE+, the results are consistent across matched and mismatched conditions: the M6 model (PESQ-high) leads in \ac{PESQ}, CSIG and COVL by a significant margin over M5 (no PESQ); M7 (PESQ-mid) yields the best POLQA and CBAK values; and M5 gets the highest \ac{segSNR}, SDR and ESTOI scores. Other metrics generally show close performance between no PESQ and PESQ-mid. For SEMamba, results differ across test sets: in the matched set, performance is generally similar across metrics, with the exception of \ac{PESQ} and composite metrics. In contrast, the mismatched set shows a clear distinction, with most metrics favoring the model without a metric loss, including the aforementioned \ac{PESQ} and composite measures. This might be explained by the fact that the metric discriminator itself is also trained on \ac{VB-DMD}, concurrently with the \ac{SE} model. The limited size and \ac{SNR} range of \ac{VB-DMD} lead to a weak generalization ability of the discriminator, which is visible in the larger gaps to real \ac{PESQ} values on EARS-WHAM.

\begin{figure}[t]
    \centering
    \includegraphics[width=\linewidth]{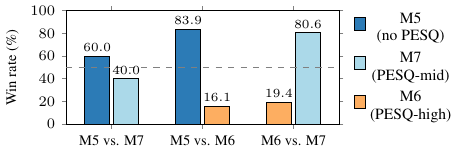}
    \vspace{-0.7cm}
    \caption{Listening experiment pairwise win rates (\%) for SB-SGMSE+.}
    \label{fig:winrate_pairwise}
    \vspace{-0.2cm}
\end{figure}

The listening experiment results reveal a statistically significant general preference for the models optimized without \ac{PESQ} loss terms. Fig.~\ref{fig:winrate} shows the win rates of each model, i.e., the percentage of trials in which the model was preferred when featured in a pairwise comparison. In the matched \ac{VB-DMD} case, SB-SGMSE+ with no PESQ and PESQ-mid have close performance, and the same holds for SEMamba with and without metric loss. The differences become clearer in the mismatched case, where the no-PESQ models gain approximately 15 percentage points across both models. In the SB-SGMSE+ analysis, the PESQ-mid model remains stable across conditions, but still behind the model without \ac{PESQ}. Fig.~\ref{fig:winrate_pairwise} displays the pairwise win rates. A Bradley-Terry model is applied to convert these into a ranking, indicating the no-PESQ model as the strongest model of the trio (0.534), followed by PESQ-mid (0.371), and PESQ-high ranking lowest (0.095). Corroborating the findings of \cite{deoliveira2024pesqetarian}, this shows that too much focus on \ac{PESQ} optimization is counterproductive to \ac{SE} performance. For the SEMamba models, a binomial test rejects the hypothesis that listeners' preference for the model without a metric loss is due to chance ($p=.030$).

Table~\ref{tab:composite_importance} shows the explained variance of each metric within each composite linear combination, following the methodology of Sec.~\ref{subsec:composite}. Please note that, although the code commonly used to compute them derives from the Matlab script published alongside \cite{loizou2013speech}, the metrics and weights in that script actually correspond to \cite{hu2006evaluation}. In the data of this experiment, \ac{PESQ} dominates the variance in all composite measures, while \ac{WSS} accounts for less than 1\% in any of them. Combined with the low variance in ESTOI, this highlights the importance of using a wider a set of metrics beyond CMGAN's evaluation script.

\begin{table}[t]
    \caption{Effective importance of individual metrics in the composite measures as implemented in \cite{hu2006evaluation}. Empirical statistics were computed over all models and both datasets.}
    \vspace{0.1cm}
    \label{tab:composite_importance}
    \centering
    \begin{adjustbox}{width=\linewidth}
    \begin{tabular}{llrllrllr}
        \toprule
        & Metric & \% Var & & Metric & \% Var & & Metric & \% Var \\
        \cmidrule(r){1-3}\cmidrule(lr){4-6}\cmidrule(l){7-9}
        \parbox[c]{2mm}{\multirow{3}{*}{\rotatebox[origin=c]{90}{CSIG}}} & \ac{PESQ} & \textbf{59.2} & \parbox[c]{2mm}{\multirow{3}{*}{\rotatebox[origin=c]{90}{CBAK}}} & \ac{PESQ} & \textbf{69.0} & \parbox[c]{2mm}{\multirow{3}{*}{\rotatebox[origin=c]{90}{COVL}}} & \ac{PESQ} & \textbf{91.1} \\
        & LLR & 40.2 & & segSNR & 30.3 & & LLR & 8.6 \\
        & WSS & 0.7 & & WSS & 0.8 & & WSS & 0.4 \\
        \bottomrule
    \end{tabular}
    \end{adjustbox}
    \vspace{-0.4cm}
\end{table}

\vspace{-0.25cm}
\section{CONCLUSION}
\label{sec:conclusion}

This paper presents a thorough analysis of two representative \ac{SE} models with and without a perceptually-motivated loss term (PESQ). Both a differentiable \ac{PESQ} and a \ac{PESQ}-surrogate GAN-style discriminator implementations were considered. \ac{SE} performance was evaluated using a varied set of instrumental metrics and found no consensus beyond \ac{PESQ} and composite metrics, which were shown to be dominated by \ac{PESQ} via a variance analysis. In a mismatched setting, the GAN-style loss term was found to be even detrimental to \ac{PESQ}. A formal listening experiment confirmed listening preference for the models without a \ac{PESQ} loss, a phenomenon more pronounced in a mismatched setting. Our work demonstrates the risks and caveats of perceptual losses in speech enhancement, even in controlled doses, and highlights the importance of a complete evaluation procedure.
\vspace{-.1cm}
\blfootnote{For this work, the HPC cluster Hummel-2 at the University of Hamburg was used, funded by the Deutsche Forschungsgemeinschaft (DFG) - 498394658. The authors would like to thank J. Berger and Rohde\&Schwarz SwissQual AG for their support with POLQA.}

\section{REFERENCES}
\label{sec:refs}

\printbibliography[heading=none]

\end{document}

%% file: acronyms.tex
\begin{acronym}
\acro{ACR}{absolute category rating}
\acro{ASR}{automatic speech recognition}
\acro{DNN}{deep neural network}
\acro{PER}{phoneme error rate}
\acro{WER}{word error rate}
\acro{SNR}{signal-to-noise ratio}
\acro{segSNR}{segmental signal-to-noise ratio}
\acro{SI-SDR}{scale invariant signal-to-distortion ratio}
\acro{STFT}{short-time Fourier transform}
\acro{PESQ}{Perceptual Evaluation of Speech Quality}
\acro{POLQA}{Perceptual Objective Listening Quality Analysis}
\acro{STOI}{short-time objective intelligibility}
\acro{DNSMOS}{deep noise suppression mean opinion score}
\acro{MOS}{mean opinion score}
\acro{LPS}{Levenshtein phoneme similarity}
\acro{CTC}{connectionist temporal classification}
\acro{SSL}{self-supervised learning}
\acro{FFT}{fast Fourier transform}
\acro{GPU}{graphics processing unit}
\acro{ROI}{region-of-interest}
\acro{EMA}{exponential moving average}
\acro{DAC}{descript audio codec}
\acro{VSR}{visual speech recognition}
\acro{EMA}{exponential moving average}
\acro{LM}{language model}
\acro{LLM}{large language model}
\acro{uLM}{unit language model}
\acro{SpeechLM}{speech language model}
\acro{VQ-VAE}{vector-quantized variational autoencoder}
\acro{VB-DMD}{VoiceBank-DEMAND}
\acro{SE}{speech enhancement}
\acro{PCC}{Pearson's correlation coefficient}
\acro{SRCC}{Spearman's rank correlation coefficient}
\acro{L2S}{lip-to-speech}
\acro{GSLM}{Generative Spoken Language Modeling}
\acro{SB}{Schr\"odinger Bridge}
\acro{VAD}{voice activity detector}
\acro{ITU}{International Telecommunication Union}
\acro{SQA}{speech quality assessment}
\acro{WSS}{weighted-slope spectral}
\acro{SCOREQ}{Speech Contrastive Regression for Quality Assessment}
\acro{GAN}{generative adversarial network}
\end{acronym}